\documentclass[twocolumn,prl]{revtex4-1}
\usepackage{amsfonts}
\usepackage[T1]{fontenc}
\usepackage{amsmath,amsbsy,amssymb,graphicx}
\usepackage{amsmath}
\usepackage{amssymb}
\usepackage{graphicx}
\usepackage{times}

\def\beginABC{\begin{subequations}}
\def\endABC{\end{subequations}}
\let\mathbf=\boldsymbol

\begin{document}

\title{{\Large Giant Skyrmions Stabilized by Dipole-Dipole Interactions}\\
{\Large in Thin Ferromagnetic Films}}
\author{Motohiko Ezawa}

\begin{abstract}
Motivated by a recent magnetization reversal experiment on a TbFeCo thin
film, we study a topological excitation in the anisotropic nonlinear sigma
model together with the Zeeman and magnetic dipole-dipole interactions.
Dipole-dipole interactions turn a ferromagnet into a frustrated spin system,
which allows a nontrivial spin texture such as a giant skyrmion. We derive
an analytic formula for the skyrmion radius. The radius is controllable by
the external magnetic field. It is intriguing that a skyrmion may have
already been observed as a magnetic domain. A salient feature is that a
single skyrmion can be created or destroyed experimentally. An analysis is
made also on skyrmions in chiral magnets.
\end{abstract}

\date{\today }
\maketitle

\affiliation{Department of Applied Physics, University of Tokyo, Hongo 7-3-1, 113-8656,
Japan }

Skyrmions, originally proposed to account for baryons in nuclear physics,
now play crucial roles in almost all branches of physics\cite{Skyrmion}. In
condensed matter physics, in particular, the notion of skyrmion has proved
to be most successful, manifesting itself in various experimental
observations. Skyrmions are solitons in a nonlinear field theory
characterized by the topological quantum number. In addition to its
topological stability, for a soliton to materialize, the Hamiltonian must\
contain extra terms that introduce scale to the system. Indeed, Skyrme
introduced a term of the fourth order in derivatives into his original model%
\cite{Skyrme}. In an instance of quantum Hall ferromagnets, the scale of a
skyrmion is determined by the competitive interplay between the Zeeman and
the Coulomb interactions\cite{Sondhi,Abolfath}. Recently, a skyrmion lattice%
\cite{Mohlbauer} as well as a single skyrmion\cite{Yu} have been observed in
chiral magnets, where the role of the Dzyaloshinskii-Moriya interaction
(DMI) has been discussed\cite{Pfleiderer,HanNagaoka}.

In this paper we propose a new mechanism of skyrmion materialization based
on the magnetic dipole-dipole interaction (DDI). Magnetic DDIs exist in all
magnetic materials. They introduce frustration into a ferromagnetic state.
Frustration between DDIs and exchange interactions could lead to rich
phenomena in magnetic materials. Indeed, the ground state of a thin
ferromagnet film has a complicated magnetic domain structure comprised of
stripes, bubbles or labyrinths\cite{Yafet,Kashuba,Ng,Bader,Portmann}. We
show that a giant skyrmion emerges in external magnetic field. The size of
the skyrmion is so large ($\sim 1\mu m$) that it may be observed simply as a
magnetic domain.

We present a skyrmion spin texture in the anisotropic nonlinear sigma model.
Deriving an analytic formula for the radius of a giant skyrmion, we find
that there exists a minimum radius. As the magnetic field increases, the
radius decreases and suddenly shrinks to zero, where a skyrmion disappears.
The skyrmion spin texture would make a quantum jump to the homogeneous
ground state, which breaks the conservation of the topological quantum
number. Furthermore, a single skyrmion can be created by destroying the
magnetic order within a tiny spot, as we soon argue. It is also pointed out
that the skyrmion excitation energy becomes negative, making a formation of
a skyrmion lattice possible, when the external field is less than a certain
critical value. A skyrmion enjoys the topological stability, as far as the
external magnetic field is not too large or not too weak.

\begin{figure}[t]
\centerline{\includegraphics[width=0.5\textwidth]{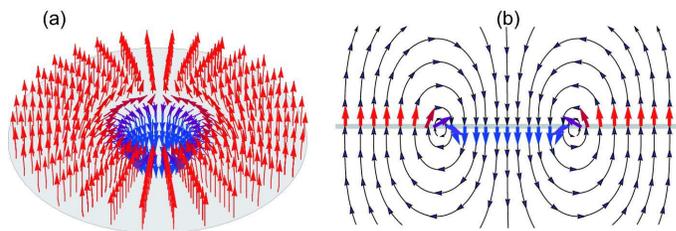}}
\caption{{}(Color online) (a) Illustration of a giant skyrmion ($\sim 1%
\protect\mu m$) in a thin ferromagnetic film. The simplest spin texture has
naturally a nontrivial Pontryagin number. It can be created by applying
femtosecond optical pulse irradiation focused on a micrometer spot and thus
destroying the magnetic order locally. (b) Illustration of magnetic flux
lines around a skyrmion due to magnetic dipoles. When the magnetic order is
destroyed locally, a new order is generated which is opposite to that of the
environs, so that the magnetic flux closes by itself as short as possible.}
\label{FigDipole}
\end{figure}

A giant skyrmion may have already been observed experimentally as a magnetic
domain in a TbFeCo thin film\cite{Ogasawara}. In this experiment the ground
state is a homogeneous spin-polarized state in a small external magnetic
field. By applying femtosecond optical pulse irradiation focused on a
micrometer spot, it is possible to destroy the magnetic order locally. Then,
the DDI generates an effective magnetic field, leading to a new magnetic
order with reversed magnetization inside the spot. The resultant spin
texture must be that of a single skyrmion, since it is the simplest and most
natural spin texture, as illustrated in Fig.\ref{FigDipole}. Thus, a
skyrmion can be created at any point from the ground state experimentally by
destroying the magnetic order within a tiny spot. This process breaks down
the continuity of the classical fields, and the conservation of the
topological quantum number is lost. A giant skyrmion may also occur in
chiral magnets, where the spin twists around the skyrmion so as to break the
chiral symmetry, as has been illustrated in Refs.\cite%
{Mohlbauer,Yu,Pfleiderer}.

Our system is a two-dimensional ferromagnetic plane in perpendicular
magnetic field. The basic Hamiltonian is the nonlinear O(3) sigma model $%
H_{J}$ with easy axis anisotropy%
\begin{equation}
H_{J}=\frac{1}{2}\Gamma \int d^{2}x[\left( \partial _{k}\mathbf{n}\right)
\left( \partial _{k}\mathbf{n}\right) -\xi ^{-2}\left( n_{z}\right) ^{2}],
\label{SigmaModel}
\end{equation}%
where $\Gamma =(1/2)zS^{2}J$ is the exchange energy\cite{Kashuba} ($z$
denotes the number of nearest neighbors, $S$ the spin per atom, $J$ the
exchange constant), $\xi $ is the single-ion anisotropy constant, and $%
\mathbf{n}=(n_{x},n_{y},n_{z})$ is a classical field of unit length. The
ground-state solutions of $H_{J}$ are $\mathbf{n}=(0,0,\pm 1)$.

The DDI between two spins is described by the term%
\begin{equation}
H_{D}=\frac{\Omega }{4\pi }\int d^{2}xd^{2}x^{\prime }\left[ \frac{\mathbf{n}%
(\mathbf{x})\cdot \mathbf{n}(\mathbf{x}^{\prime })}{r^{3}}-\frac{3[\mathbf{n}%
(\mathbf{x})\cdot \mathbf{r}][\mathbf{n}(\mathbf{x}^{\prime })\cdot \mathbf{r%
}]}{r^{5}}\right]  \label{DipolInter}
\end{equation}%
with $\mathbf{r}=\mathbf{x}-\mathbf{x}^{\prime }$ and $r=|\mathbf{r}|$,
where $\Omega =NS^{2}g^{2}\mu _{B}^{2}\mu _{0}/a^{4}$ is the DDI strength%
\cite{Kashuba} ($N$ denotes the number of the layers, $g$ the Land\'{e}
factor, $\mu _{B}$ the Bohr magneton, $a$ the lattice constant).

We apply the magnetic field $h$ perpendicular to the plane,%
\begin{equation}
H_{Z}=-\Delta _{Z}\int \frac{d^{2}x}{a^{2}}\,n_{z}(\mathbf{x}),
\label{ZeemaInter}
\end{equation}%
with the Zeeman energy $\Delta _{Z}=Sg\mu _{B}\mu _{0}h$, so that the ground
state is the spin-polarized homogeneous state, $\mathbf{n}=(0,0,1)$.

The ordinary ferromagnet is described by the Hamiltonian $%
H=H_{J}+H_{D}+H_{Z} $. On the other hand, the chiral magnet is described by
the Hamiltonian $H=H_{J}+H_{D}+H_{Z}+H_{\text{DM}}$, where%
\begin{equation}
H_{\text{DM}}=D\mathbf{n}(\mathbf{x})\cdot \left( \nabla \times \mathbf{n}(%
\mathbf{x})\right)  \label{DMI}
\end{equation}%
is the DMI term. It breaks the chiral symmetry explicitly.

The use of a continuum approximation and of classical fields to represent
the spins is justified as far as we analyze phenomena whose characteristic
wavelength is much larger than the lattice constant.

We first analyze the ordinary ferromagnet and then the chiral magnet. We
start with the study of the ground state in weak external magnetic field. It
is well known that the DDI makes the ground state to have an alternating
up-down stripe-domain structure. Our results are consistent with those in
literature\cite{Yafet,Kashuba,Ng,Bader,Portmann} but we also present some
new formulas. The basic object in our analysis is the one-dimensional
solution of the anisotropic nonlinear sigma model (\ref{SigmaModel})
independent of the $y$ coordinate. It has two types of solutions; a kink
localized in the $x$-axis and a soliton lattice periodic in the $x$-axis.

The kink solution of the nonlinear sigma model (\ref{SigmaModel}) is given
by $n_{x}(x)=\sqrt{1-\sigma ^{2}(x)}$, $n_{y}=0$, $n_{z}(x)=\sigma (x)$ with%
\begin{equation}
\sigma (x)=\pm \tanh [({x-x}_{0})/\xi ].  \label{KinkBasic}
\end{equation}%
It describes a domain wall with thickness $\xi $ lying along the $y$-axis
and located at $x=x_{0}$, and separates the two field configurations $%
\mathbf{n}=(0,0,1)$ and $\mathbf{n}=(0,0,-1)$. The exchange energy is $E_{%
\text{kink}}^{J}=2\Gamma L_{y}/\xi $, where $L_{y}$ is the length of the
kink along the $y$-axis. The DDI energy (\ref{DipolInter}) is found to be
negative and divergent for a fixed value of $L_{y}$, indicating the ground
state is a condensed phase of domain walls in the absence of the Zeeman
effect.

\begin{figure}[t]
\centerline{\includegraphics[width=0.46\textwidth]{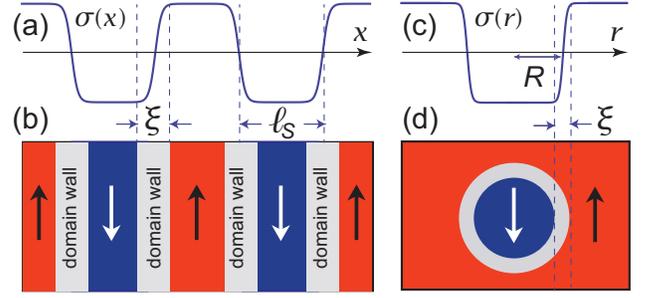}}
\caption{{}(Color online) (a) Illustration of a soliton-lattice solution $%
\protect\sigma (x)$ described by the Jacobian elliptic function. (b)
Illustration of the ground-state structure with an alternating up-down
stripe-domains separated by domain walls. (c) Illustration of a skyrmion
field configuration $\protect\sigma (x)$ described by (\protect\ref%
{SkyrmField}) with (\protect\ref{SkyrmSigma}). (d) Illustration of a
skyrmion spin texture with up and down domains separated by a circular
domain wall with thickness $\protect\xi $. }
\label{FigSL}
\end{figure}

Such a phase is described by the periodic solution,%
\begin{equation}
\sigma (x)=\text{sn}\left( x/\kappa \xi ,\kappa ^{2}\right) ,
\label{DomaiWall}
\end{equation}%
in terms of the Jacobian elliptic function with $0<\kappa ^{2}\leq 1$: See
Fig.\ref{FigSL}(a). The periodicity of $\sigma (x)$ is $2\ell _{S}$ with%
\begin{equation}
\ell _{S}\equiv 2\kappa \xi K(\kappa ^{2}),
\end{equation}%
where $K(\kappa ^{2})$ is the complete elliptic integral of the first kind.
The soliton-lattice solution (\ref{DomaiWall}) implies that the ground state
has an alternating up-down stripe-domain structure separated by domain
walls, as illustrated in Fig.\ref{FigSL}(b). The width of one stripe is
given by $\ell _{S}$.

Substituting (\ref{DomaiWall}) into the Hamiltonian (\ref{SigmaModel}) we
obtain the energy of the soliton lattice to be%
\begin{equation}
E_{\text{SL}}^{J}=\frac{\Gamma }{2\xi ^{2}}L_{x}L_{y}\left[ \frac{2}{\kappa
^{2}}\frac{E(\kappa ^{2})}{K(\kappa ^{2})}-\frac{1}{\kappa ^{2}}+1\right]
\simeq (L_{x}/\ell _{S})E_{\text{kink}}^{J},
\end{equation}%
where $E(\kappa ^{2})$ is the complete elliptic integral of the second kind.
The number of solitons is given by $L_{x}/\ell _{S}$. The last equality
holds in the regime where $\kappa \approx 1$ or $\ell _{S}/\xi \gg 1$.

We are able to estimate the DDI energy for $\ell _{S}/\xi \gg 1$ as 
\begin{equation}
E_{\text{SL}}^{D}=-\frac{\Omega }{2\pi }\frac{L_{x}L_{y}}{\ell _{S}}\ln 
\frac{\ell _{S}}{ed_{F}},
\end{equation}%
where $d_{F}$ is the thickness of the film. The Zeeman interaction does not
make any contribution.

The total energy is%
\begin{equation}
E_{\text{SL}}=\frac{L_{x}L_{y}}{\ell _{S}}\left( \frac{2\Gamma }{\xi }-\frac{%
\Omega }{2\pi }\ln \frac{\ell _{S}}{ed_{F}}\right) .
\end{equation}%
Minimizing this with respect to $\ell _{S}$, we obtain%
\begin{equation}
\ell _{S}=d_{F}\exp \left( 2+4\pi \Gamma /\xi \Omega \right) .
\label{WidthStrip}
\end{equation}%
The ground-state energy is%
\begin{equation}
E_{\text{SL}}=-(L_{x}L_{y}\Omega /2\pi d_{F}e^{2})e^{-4\pi \Gamma /\Omega
\xi }.  \label{EnergStrip}
\end{equation}%
The stripe-domain structure is not robust and may be deformed into a
labyrinth structure. The formula (\ref{WidthStrip}) gives the order of the
width of a stripe segment in this case .

On the other hand, in strong external magnetic field we expect the
homogeneous state, which has only the Zeeman energy, $E_{\text{homo}%
}=-\left( L_{x}L_{y}/a^{2}\right) \Delta _{Z}$. By comparing this with (\ref%
{EnergStrip}), there exists the critical magnetic field $h_{c}$ with the
associated Zeeman energy being%
\begin{equation}
\Delta _{Z}^{c}=(a^{2}\Omega /2\pi d_{F}e^{2})e^{-4\pi \Gamma /\Omega \xi }.
\label{CritiZeema}
\end{equation}%
The ferromagnet phase appears for $\Delta _{Z}>\Delta _{Z}^{c}$, or $h>h_{c}$%
.

We proceed to analyze a skyrmion spin texture in the ferromagnet phase. In
the regime where the classical-field approximation is valid, there exists
the topologically conserved charge, that is the Pontryagin number, 
\begin{equation}
Q_{\text{sky}}=-{\frac{1}{8\pi }}\sum_{ij}\int \!d^{2}x\,\varepsilon _{ij}%
\mathbf{n}(\mathbf{x})\cdot \left( \partial _{i}\mathbf{n}(\mathbf{x})\times
\partial _{j}\mathbf{n}(\mathbf{x})\right) ,  \label{PontrNumbe}
\end{equation}%
where $i,j$ run over $x,y$ with $\varepsilon _{ij}$ being the completely
antisymmetric tensor. A spin texture possessing nonzero $Q_{\text{sky}}$ is
a skyrmion by definition.

We consider the spin texture given by%
\begin{align}
n_{x}& =-\sqrt{1-\sigma ^{2}(r)}\cos (\theta +\theta _{0}),  \notag \\
n_{y}& =-\sqrt{1-\sigma ^{2}(r)}\sin \left( \theta +\theta _{0}\right)
,\quad n_{z}=\sigma (r)  \label{SkyrmField}
\end{align}%
in the cylindrical coordinate, where $\theta _{0}$ is a constant parameter
representing a zero-energy mode in the nonlinear sigma model $H_{J}$, while $%
\sigma (r)$ is a trial function. The boundary conditions are $\sigma
(r)\rightarrow 1$ as $r\rightarrow \infty $, and $\sigma (r)=-1$ at $r=0$.
The first condition implies this describes a soliton in the ground state
with $\sigma (r)=1$. The second condition is required to remove the
multivalueness of the field $\mathbf{n}(\mathbf{x})$ at the center.
Substituting (\ref{SkyrmField}) into (\ref{PontrNumbe}), we find that $Q_{%
\text{sky}}=1$ for any $\theta _{0}$, where the use of the boundary
conditions on $\sigma (r)$ is made. Hence, the spin texture (\ref{SkyrmField}%
) has one unit charge, and hence it is a skyrmion. Its stability is
guaranteed topologically.

The chiral symmetric configuration is given by the choice of $\theta _{0}=0$%
, which minimizes the DDI energy. This is the simplest and most natural
field configuration in the nonchiral magnet, as illustrated in Fig.\ref%
{FigDipole}.

We consider a domain with radius $R$ separated by a domain wall, where $%
\sigma (r)\approx -1$ for $r<R$ and $\sigma (r)\approx 1$ for $r>R$. The
wall must have thickness $\xi $ due to the anisotropy term in (\ref%
{SigmaModel}). Such a domain structure is well described by [Fig.\ref{FigSL}%
(c)] 
\begin{equation}
\sigma (r)=\tanh \left[ (R/\xi )\log (r/R)\right] ,  \label{SkyrmSigma}
\end{equation}%
which satisfies the boundary conditions at $r=0$ and $r\rightarrow \infty $.
Near the domain wall, it behaves as%
\begin{equation}
\sigma (r)=\tanh [{(r-R)/\xi }]\quad \text{at}\quad r\approx R,
\end{equation}%
provided $R\gg \xi $, as agrees with the domain-wall solution (\ref%
{DomaiWall}).

Provided $R\gg \xi $, $\sigma (r)\approx \pm 1$ except for the domain-wall
region, and the exchange energy is estimated as 
\begin{equation}
E_{\text{sky}}^{J}=4\pi \Gamma R/\xi .  \label{EnergSkyJ}
\end{equation}%
The skyrmion is not physical as it stands, though it carries the topological
number, since the energy can be made zero by letting $R\rightarrow 0$ in (%
\ref{EnergSkyJ}). It is made physical by the DDI in addition to the Zeeman
effect, as we now show.

\begin{figure}[t]
\centerline{\includegraphics[width=0.48\textwidth]{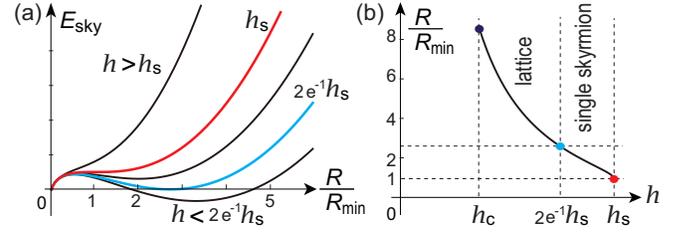}}
\caption{{}(Color online) (a) The skyrmion size $R$ is determined by
minimizing the energy $E_{\text{sky}}(R)$. Various curves are for different
magnetic field $h$. Skyrmion is stable only for $h<h_{s}$. (b) The skyrmion
size $R$ decreases as the field $h$ increases. A skyrmion lattice may appear
for $h_{c}<h<2e^{-1}h_{s}$.}
\label{FigLambert}
\end{figure}

The DDI energy is estimated by substituting the skyrmion configuration (\ref%
{SkyrmSigma}) into the DDI Hamiltonian (\ref{DipolInter}), 
\begin{equation}
E_{\text{sky}}^{D}=-\Omega \lbrack R\ln (R/d_{F})-R].
\end{equation}%
The total excitation energy is given by%
\begin{equation}
E_{\text{sky}}(R)=\frac{4\pi \Gamma R}{\xi }-\Omega \lbrack R\ln \frac{R}{%
d_{F}}-R]+\pi \frac{R^{2}}{a^{2}}\Delta _{Z},  \label{SkyrmEnerg}
\end{equation}%
where the last term is the Zeeman energy. We have illustrated $E_{\text{sky}%
}(R)$ for typical values of $\Delta _{Z}$ in Fig.\ref{FigLambert}. The
variation of (\ref{SkyrmEnerg}) with respect to $R$ yields,%
\begin{equation}
\frac{4\pi \Gamma }{\xi }-\Omega \ln \frac{R}{d_{F}}+\frac{2\pi \Delta _{Z}}{%
a^{2}}R=0.  \label{Emin}
\end{equation}%
Solving this for $R$ we find%
\begin{equation}
R=-\frac{a^{2}\Omega }{2\pi \Delta _{Z}}W_{-1}\left( \frac{\Delta _{Z}}{%
e\Delta _{Z}^{s}}\right) ,  \label{SkyrmRadiu}
\end{equation}%
where $\Delta _{Z}^{s}=e(a^{2}\Omega /2\pi d_{F}e^{2})\exp \left( -4\pi
\Gamma /\Omega \xi \right) $, and $W_{-1}(z)$ is the Lambert function. We
may also solve (\ref{Emin}) for $\Omega $,%
\begin{equation}
\Omega =\frac{4\pi \Gamma /\xi +2\pi \Delta _{Z}R/a^{2}}{\log (R/d_{F})}.
\label{StepA}
\end{equation}%
This is interpreted as the formula to determine the effective DDI strength
from the size of a skyrmion. A comment is in order: It follows that $\Delta
_{Z}^{s}=e\Delta _{Z}^{c}$ with (\ref{CritiZeema}). This is a relation
peculiar to the nonchiral magnet.

The Lambert function $W_{-1}(z)$ has a real value only for $z>-1/e$, as
implies%
\begin{equation}
\Delta _{Z}<\Delta _{Z}^{s},\quad \text{or\quad }h<h_{s}.
\end{equation}%
Namely, the skyrmion is stable only when the magnetic field is less than the
critical one $h_{s}$. The minimum radius is%
\begin{equation}
R_{\text{min}}=a^{2}\Omega /\left( 2\pi \Delta _{Z}^{s}\right) =\ell _{S}/e,
\label{Rmin}
\end{equation}%
where $\ell _{S}$ is the width of one stripe in the absence of the magnetic
field and given by (\ref{WidthStrip}).

When $h$ exceeds $h_{s}$, the radius shrinks to zero and the skyrmion
disappears, which is beyond the classical-field regime of the spin. What
actually happens would be that the spin direction at the skyrmion center
makes a quantum jump to that of the homogeneous ground state.

The skyrmion energy is less than the ground-state energy for $h<2e^{-1}h_{s}$%
, as illustrated in Fig.\ref{FigLambert}. Skyrmions would condense and form
a lattice to make a new ground state when the sample is cooled from high
temperature in the magnetic field $h<2e^{-1}h_{s}$. On the other hand, once
a skyrmion is created in the ferromagnetic ground state ($h>2e^{-1}h_{s}$)
and then the magnetic field is decreased at low temperature, it remains to
be a stable object even for $h<2e^{-1}h_{s}$.

We next discuss a skyrmion in chiral magnets such as MnSi and FeCoSi thin
films. The Hamiltonian is given by $H=H_{J}+H_{D}+H_{Z}+H_{\text{DM}}$,
where $H_{\text{DM}}$ is the DMI term (\ref{DMI}). The system is in the
ferromagnetic phase beyond a certain critical magnetic field $h_{c}$. The
skyrmion spin texture is given by (\ref{SkyrmField}), where the zero-energy
mode $\theta _{0}$ of the nonlinear sigma model $H_{J}$ is fixed by
minimizing the total energy. We find $\theta _{0}=-\pi /2$ due to the DMI
term. Namely, the spin twists around a skyrmion as in the illustrations
given in Refs.\cite{Mohlbauer,Yu,Pfleiderer}. All the rest analysis can be
carried out almost as it was. Indeed, the function $\sigma (r)$ is given by (%
\ref{SkyrmSigma}). Provided $R\gg \xi $, the DMI energy is calculated to be 
\begin{equation}
E_{\text{sky}}^{\text{DM}}=-2\pi DR.
\end{equation}%
There exists a small increase of the DDI energy due to the twisting of spins
inside the domain wall, which we neglect. As a result the net effect is to
renormalize the exchange energy as 
\begin{equation}
\Gamma \rightarrow \Gamma _{\text{DM}}\equiv \Gamma -\frac{1}{2}\xi D
\end{equation}%
in the skyrmion excitation energy (\ref{SkyrmEnerg}). Then, we can derive (%
\ref{Emin})$\sim $(\ref{Rmin}) just as they are simply by replacing $\Gamma $
with $\Gamma _{\text{DM}}$. A giant skyrmion stabilized by the DDI may also
occur in the chiral magnet provided $\Gamma >\frac{1}{2}\xi D$. The
essential role of the DMI is to twist the spin texture around a skyrmion.

The present work was motivated by a recent magnetization reversal experiment%
\cite{Ogasawara} on a TbFeCo thin film. It is not certain if our analysis
can be applicable to this material as it stands, because TbFeCo is amorphous
and what is realized is a ferrimagnet. Nevertheless, it is tempting to make
a comparison of our analytical results to the experimental results by
identifying the observed magnetization reversal domain as a skyrmion. Sample
parameters\cite{Ogasawara,Rahman} are the lattice constant $a=0.3$ [nm], the
layer thickness $d_{F}=35$ [nm] and the exchange energy $\Gamma =3.5\times
10^{-20}$[J]. The experiment was carried out in the magnetic field at $170$%
[Oe], where the radius of the observed skyrmion is about $R=1$ [$\mu $m] and
the thickness of the domain wall is about $\xi =25$[nm]. We then use (\ref%
{StepA}) to determines the effective DDI strength, $\Omega =8.4\times
10^{-12}\left[ \text{J/m}\right] $. This is a good agreement with $\Omega
=7.0\times 10^{-12}\left[ \text{J/m}\right] $ for TbFeCo, which is evaluated
from its definition given below (\ref{DipolInter}). We predict the minimum
skyrmion radius to be $R_{\text{min}}=0.7\left[ \mu \text{m}\right] $. Using
these parameters, the stripe width becomes of the order of $\ell _{S}=1.9$ [$%
\mu $m] in the absence of the magnetic field. On the other hand, $h_{s}=190$
[Oe], beyond which a skyrmion is expected to disappear.

I am deeply indebted to Y. Tokura and T. Ogasawara for illuminating
discussions and for informing me as to experimental details. I am very much
grateful to N. Nagaosa for fruitful discussions on the subject. This work
was supported in part by Grants-in-Aid for Scientific Research from the
Ministry of Education, Science, Sports and Culture No. 22740196 and 21244053.


\begin{thebibliography}{99}
\bibitem{Skyrmion} G.E. Brown and M. Rho (eds.), "The Multifaced Skyrmions",
World Scientific, Singapore (2010).

\bibitem{Skyrme} T. H. R. Skyrme, Proc. Roy. Soc. (London) A \textbf{260},
127 (1961); Nuc. Phys. \textbf{31}, 556 (1962).

\bibitem{Sondhi} S. L. Sondhi, \textit{et al}. Phys. Rev. B \textbf{47},
16419 (1993).

\bibitem{Abolfath} M. Abolfath, \textit{et al}. Phys. Rev. B \textbf{56,}
6795 (1997).

\bibitem{Mohlbauer} Mohlbauer et al., Science 323, 915 (2009); Monzer et al.
Phys. Rev. B 81, 041203 (2010).

\bibitem{Yu} X. Z. Yu, Y. Onose, N. Kanazawa, J. H. Park, J. H. Han, Y.
Matsui, N. Nagaosa and Y. Tokura Nature, \textbf{465}, 901 (2010).

\bibitem{Pfleiderer} C. Pfleiderer and Achim Rosch, Nature \textbf{465}, 880
(2010).

\bibitem{HanNagaoka} J. H. Han, J. Zang, Z. Yang, J.-H. Park and N. Nagaosa,
Phys. Rev. B 82, 094429 (2010).

\bibitem{Yafet} Y. Yafet and E.M. Gyogy, Phys. Rev. B, \textbf{38}, 9145
(1988).

\bibitem{Kashuba} A.B. Kashuba and V.L. Pokrovsky, Phys. Rev. B, \textbf{48}%
, 10335 (1993).

\bibitem{Ng} K-O. Ng and D. Vanderbilt, Phys. Rev. B, \textbf{52}, 2177
(1995).

\bibitem{Bader} S.D. Bader, Rev. Mod. Phys., \textbf{78}, 1 (2006).

\bibitem{Portmann} O. Portmann, A. Vaterlaus and D. Pescia, Nature, \textbf{%
422}, 701\ (2003).

\bibitem{Ogasawara} T. Ogasawara, N. Iwata, Y. Murakami, H. Okamoto and Y.
Tokura, Appl. Phys. Lett. \textbf{94}, 162507 (2009).

\bibitem{Rahman} M. T. Rahman, \textit{et al}. J. Appl. Phys. \textbf{97},
10C515 (2005).
\end{thebibliography}
\end{document}